\documentstyle[pre,aps]{revtex}
\input{psfig.sty}
\input{epsf}
\draft
\begin{document}

\title{Plasticity and learning in a network of coupled phase oscillators}
\author{Philip Seliger, Stephen C. Young, and Lev S. Tsimring}
\address{
Institute for Nonlinear Science, University of California,  
San Diego, La Jolla, CA 92093-0402 }
\date{\today}

\maketitle

\begin{abstract} 
A generalized Kuramoto model of coupled phase oscillators with slowly
varying coupling matrix is studied. The dynamics of the coupling
coefficients is driven by the phase difference of pairs of oscillators in
such a way that the coupling strengthens for synchronized oscillators
and weakens for non-synchronized pairs. The system possesses a family of
stable solutions corresponding to synchronized clusters of different
sizes.  A particular cluster can be formed by applying external driving at a
given frequency to a group of oscillators.  Once established, the
synchronized state is robust against noise and small variations in
natural frequencies.  The phase differences between oscillators within
the synchronized cluster can be used for information storage and
retrieval.  
\end{abstract}

\pacs{PACS: 05.45.Xt, 07.05.Mh, 87.18.Sn, 87.19.La}


\section{Introduction}
The mechanisms of adaptation and learning in natural and artificial
systems are the subject of paramount interest in neuroscience.  Most of
the experimental evidence point to the synaptic modification as the
physiological basis of the long-term memory \cite{tsien,abbott}. Hebb
\cite{hebb} was the first who suggested that the synaptic coupling
between two neurons is enhanced if both neurons are simultaneously
active. Recent experimental data suggest that the synaptic connections
are strengthened (long-term potentiation) or weakened (long-term
depression) depending on the precise relative timing between
pre-synaptic and post-synaptic spikes\cite{abbott,markram97,poo}.

Many current theoretical studies of learning in neural systems are based on
the quasi-static Hopfield model\cite{hopfield} utilizing McCulloch-Pitts
neuronal units or simple ``integrate-and-fire'' models of neurons
coupled via synapses which accumulate the pre-synaptic activity
(spikes from other neurons) into a growing membrane potential of a
neuron\cite{tuckwell}. The latter generates its own spike (``fires")
once the membrane potential exceeds a certain threshold. The synaptic
strength is in fact not fixed, but changes in response to external
stimuli and/or inter-neural dynamics. Various models based on this
general picture \cite{roberts00,rabinovich}, have been shown to yield 
differential Hebbian learning rules.

A complimentary approach to modeling the neuronal activity involves
replacing neurons by periodic oscillators. Indeed, rhythmic activity
plays important role in many neuronal systems and functions, including
central pattern generators, visual and olfactory systems,
etc.\cite{gray}. Neural networks can often be described as networks of
coupled oscillators, so the role of relative spike timing is played by
the phases of individual oscillators (see, for example,
\cite{sompolinsky,aoyagi,hop96,hop00}). Close to the Hopf bifurcation, the
dynamics of the array of coupled oscillators can be described by the
Landau-Stewart equations for complex amplitudes of the oscillators.
Furthermore, one can reduce the dynamics of the oscillators to the pure
phase dynamics by assuming that the coupling is weak, and the amplitude
of oscillations is slaved to the phase variations. In
fact, the phase dynamics description may still be valid even away from
the Hopf bifurcation point, where the Landau-Stewart description is not
applicable.  This idea was first put forth by Winfree \cite{winfree} and
later developed by Kuramoto\cite{kuramoto} and others \cite{kur_model}.
The original Kuramoto model, in which the coupling coefficients were chosen
fixed and equal, allowed for the elegant analytical treatment within the
mean field approximation. It was shown that the system of globally
coupled oscillators with non-identical frequencies exhibits a
second-order phase transition as the coupling strength is increased
above some critical value.

In subsequent work, more complex coupling functions were
considered\cite{hop00,strogatz1,hop99}. It was shown that depending on
the choice of the connectivity matrix, synchronized states with
non-trivial phase relationship among the oscillators can emerge.
Networks of coupled phase oscillators with appropriately tuned coupling
matrix were shown to have neurocomputing properties similar to those of classical
Hopfield networks, with an important distinction that memorized patterns
are stored in the form of relative phases of synchronized oscillators
rather than in static equilibria. 


In those studies, the connectivity matrix, however complicated, was
imposed externally to achieve the desired network dynamics.  In
fact, in order to achieve certain behavior (for example, to memorize an
image), in a system of $N$ neurons, $N^2$ coupling coefficients have to be
specified.  This would present significant difficulties in operating
such a system at large $N$.  In biological neural networks this task is
not assigned to any central control unit, but is performed in a
distributed manner via the mechanism of synaptic plasticity, i.e.  long
term potentiation or depression of synapses in response to
the dynamics of pre-synaptic and post-synaptic neurons
\cite{abbott,roberts00}.
External stimuli directly affect only the dynamics of individual
neurons and not their connections, while the synaptic plasticity is a
result of the inter-neuronal interactions.  This evolution of
connectivity matrix have been introduced in Ref.\cite{hop96} for the
Landau-Stewart model of coupled oscillators, however no analysis of the
joint dynamics of the network of oscillators {\em and} the connectivity
matrix has been presented.  In this paper, we study the simplest
possible model of this process, based on the generalization of the
Kuramoto model. In particular, we assume that the coupling coefficient
for a link between two oscillators is a slow function of the phase
difference between the two oscillators.

The paper is organized as follows. In Section \ref{model} we introduce
the generalized Kuramoto model with additional equations describing the
slow evolution of the coupling matrix. In Section \ref{two}, a simple
case of two coupled oscillators is investigated.
We show that for large enough separation of time scales of fast and
slow motions, the system is bistable: depending on initial conditions,
oscillators can be either synchronized or non-synchronized.
In Sections \ref{many},\ref{stab}, we study the existence and stability of the
family of synchronized cluster solutions in the continuum limit
$N\to\infty$. In Section \ref{memory}, we show that the generalized Kuramoto 
system can serve as a very simple model of learning and memory.
Information is stored within the synchronized cluster in the form of
phase differences between oscillators. Section \ref{conclusion}
summarizes our conclusions.

\section{Model}
\label{model}
Let us consider the dynamics of 
an ensemble of $N$ coupled phase oscillators
\begin{equation}
\dot{\phi}_i=\omega_i-\frac{1}{N}\sum_{j=1}^{N}K_{ij}F(\phi_i-\phi_j).
\label{phase_eq0}
\end{equation}
Here $\omega_i$ are natural frequencies and $\phi_i$ are phases of
individual oscillators, $F(\phi)$ is a $2\pi$-periodic coupling function, 
and $K_{ij}$ is the $N\times N$ matrix of coupling coefficients.

In order to include the mechanism of adaptation into model
(\ref{phase_eq0}), we assume that an element of the  coupling matrix
describing interaction between two
oscillators, $i$ and $j$, is controlled by the following equation,
\begin{equation}
\dot{K}_{ij}=\epsilon(G(\phi_i-\phi_j)-K_{ij}),
\label{k_eq0}
\end{equation}
where $G(\phi)$ is a $2\pi$-periodic function of its argument. 
The particular form of the coupling function $F(\phi)$ and adaptation function
$G(\phi)$ can be derived from the underlying equations describing the
dynamics of oscillators through the reduction to the phase description.
Following Kuramoto \cite{kuramoto}, we choose the simplest possible
periodic function $F(\phi)=\sin\phi$. Furthermore, we choose the
adaptation function $G(\phi)=\alpha\cos\phi$. This choice implies that
the coupling coefficient grows fastest for two oscillators which are
in-phase and decays fastest for out-of-phase oscillators. There are 
certain indication in the biological literature (see
Ref.\cite{abbott}) that
the maximal LTP and LTD correspond to small positive and negative phase 
shifts between pre-synaptic
and post-synaptic neuronal oscillations, but we are going to ignore this
complication in this model for the sake of simplicity. With the chosen
functions, our model becomes
\begin{eqnarray}
\dot{\phi}_i&=&\omega_i-\frac{1}{N}\sum_{j=1}^{N}K_{ij}\sin(\phi_i-\phi_j),
\label{phase_eq} \\
\dot{K}_{ij}&=&\epsilon(\alpha\cos(\phi_i-\phi_j)-K_{ij}).
\label{k_eq}   
\end{eqnarray}

For small $\epsilon$, the dynamics of the coupling coefficients is slow,
and one can expect that for two oscillators which are
non-synchronized, the driving term $\alpha\cos(\phi_i-\phi_j)$ is
oscillating around zero, and the resultant coupling coefficient is 
also oscillating around zero and is small $O(\epsilon)$.
On the other hand,
a pair of synchronized oscillators produces a constant non-zero driving
for the corresponding coupling coefficient, and therefore for large enough $\alpha$, the
connection between the two will be strengthened. Thus, we anticipate
a multi-stable behavior, when a group of oscillators can either be stably
synchronized or non-synchronized depending on the initial conditions for
their coupling coefficient. To illustrate this point, we first consider a
simple case of two oscillators symmetrically coupled through a single link.

\section{Two coupled oscillators}
\label{two}
In this case model (\ref{phase_eq}),(\ref{k_eq}) reduces to a set
of two coupled equations,
\begin{eqnarray}
\dot{\phi}&=&\Delta\omega-K\sin\phi,
\nonumber\\
\dot{K}&=&\epsilon (\alpha\cos\phi-K).
\label{two_osc}
\end{eqnarray}
Here $\phi=\phi_1-\phi_2$, $K=K_{12}=K_{21}$, and
$\Delta\omega=\omega_1-\omega_2$. By introducing new variables 
$\tilde{t}=\Delta\omega t, \tilde{K}=K/\Delta\omega, \tilde{\alpha}=\alpha/\Delta\omega,
\tilde{\epsilon}=\epsilon/\Delta\omega$, system (\ref{two_osc})
is simplified to 
\begin{eqnarray}
	\dot{\phi}&=&1-\tilde{K}\sin\phi,
	\nonumber\\
	\dot{\tilde{K}}&=&\tilde{\epsilon}
(\tilde{\alpha}\cos\phi-\tilde{K}).
	\label{two_osc1}
\end{eqnarray}

\begin{figure}
\centerline{ \psfig{figure=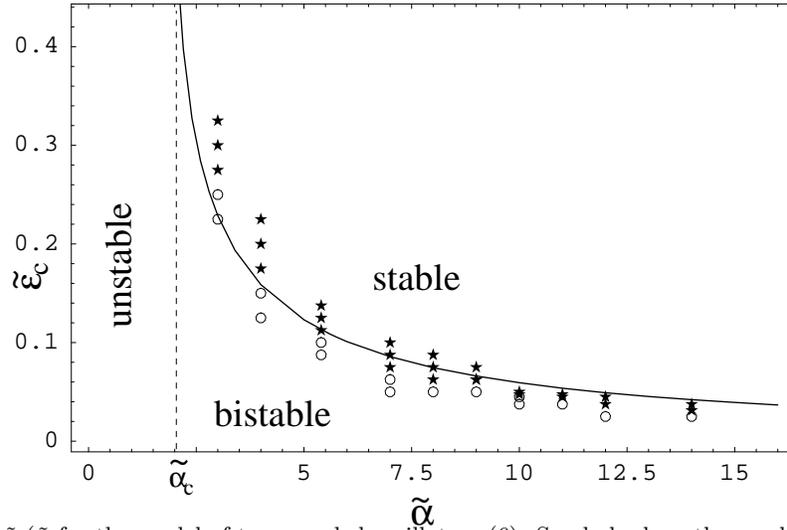,height=3.in} }
\caption{Phase diagram $\tilde{\epsilon}_c(\tilde{\alpha}$ for the model
of two
coupled oscillators (\protect\ref{two_osc1}). Symbols show the results
of numerical simulations with 50 uniformly distributed
oscillators. Circles
correspond to the non-synchronized regime, and stars indicate an onset of
synchronization of pairs of oscillators, which usually
precedes spontaneous cluster formation.  }
\label{alpha_eps}
\end{figure}

\begin{figure}
\centerline{ \psfig{figure=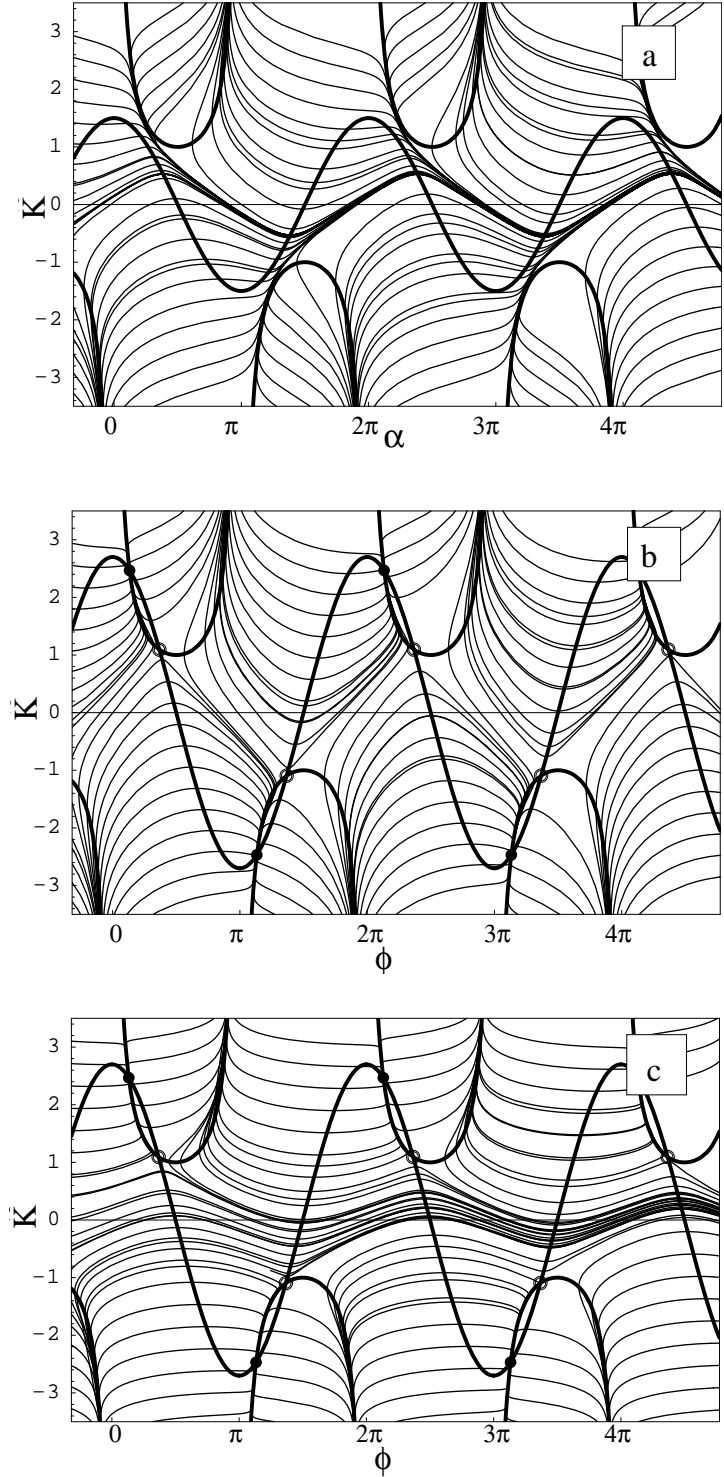,height=8.in} }   
\caption{Phase plane of Eqs.(\protect\ref{two_osc1}): {\em a} - for
$\tilde{\alpha} < \tilde{\alpha}_c$ there are no fixed points corresponding to
synchronized regime; 
{\em b} - for $\tilde{\alpha} = 2.7$ and $\tilde{\epsilon} = .33$  
all initial conditions lead to stable fixed points; 
{\em c} - for $\tilde{\alpha} = 2.7$ and $\tilde{\epsilon} = .1$ unbounded 
solutions coexist with fixed points.}
                \label{phase_plane}
\end{figure}

The nullclines of this system $\tilde{K}=\tilde{\alpha}\cos\phi$ and
$\tilde{K}=1/\sin\phi$ intersect at large enough
$\tilde{\alpha}>\tilde{\alpha}_c = 2$.  In this case, within each period of the relative
phase $\phi$, there are four intersections corresponding to two stable and
two unstable fixed points (see Figures \ref{phase_plane},b,c. 
For small
$\tilde{\epsilon}$, the dynamics of the system can be separated into fast and
slow motions. If initial value $\tilde{K}_0> 1$, the phase variable $\phi$
rapidly approaches the quasi-static value $\arcsin\tilde{K}_0^{-1}$
corresponding to the nullcline, without any significant change of $K$.
Then, depending on whether $\tilde{K}_0$ is greater or smaller than the value
$\tilde{K}_u$ corresponding to the unstable fixed point, the solution either
slowly approaches the stable fixed point $\tilde{K}_s$ (synchronized regime), or
it reaches the minimum of the nullcline and then branches off to
infinity along the fast trajectory. The same occurs if $\tilde{K}_0<1$.
This corresponds to the regime of non-synchronized oscillations. For
larger values of $\tilde{\epsilon}$, the dynamics can no longer be reduced to
fast and slow motions. Moreover, for any fixed
$\tilde{\alpha}>\tilde{\alpha}_c$, there
exists some critical value of $\tilde{\epsilon}_c$ such that at
$\tilde{\epsilon}>\tilde{\epsilon}_c$ there are no unbounded solutions corresponding to
the non-synchronized motion.  On the other hand, if
$\tilde{\alpha}<\tilde{\alpha}_c$,
there are no fixed points corresponding to the synchronized solutions, 
and the oscillators are desynchronized for any values of $\epsilon$. 
Figure \ref{phase_plane}b,c shows the structure 
of the phase plane for
system (\ref{two_osc1}) for $\tilde{\epsilon} > \tilde{\epsilon}_c$ and
$\tilde{\epsilon} < 
\tilde{\epsilon}_c$, respectively. Figure  
\ref{alpha_eps} shows the dependence of $\tilde{\epsilon}_c$ vs.
$\tilde{\alpha}$. As 
expected, $\tilde{\epsilon}_c$ diverges as
$\tilde{\alpha}\to\tilde{\alpha}_c+$. This line
together with vertical line $\tilde{\alpha}=\tilde{\alpha}_c$ divides the parameter plane 
into three regions. For $\tilde{\alpha}<\tilde{\alpha}_c$, there can only be
non-synchronized solutions (Fig.\ref{phase_plane},a), for
$\tilde{\alpha}>\tilde{\alpha}_c,
 \tilde{\epsilon}>\tilde{\epsilon}_c$ there are only synchronized solutions 
(Fig.\ref{phase_plane},b), and for $\tilde{\alpha}>\tilde{\alpha}_c,
\tilde{\epsilon}<\tilde{\epsilon}_c$, there is a bistable state when both desynchronized
and synchronized regimes are stable (Fig.\ref{phase_plane},c).  

The results of this section can be applied to a more generic case of
many coupled oscillators. Indeed, as $\epsilon$ and/or $\alpha$ is increased, 
the spontaneous clustering usually begins from synchronization of pairs
of neighboring oscillators. For this process, the coupling between these
oscillators and other oscillators can be neglected. Symbols in Fig.
\ref{alpha_eps} show the results of numerical experiments with a
population of 50 oscillators uniformly distributed within a range
$\Delta\Omega=1$ starting from small randomized initial coupling
$K_{ij}$. The bifurcation to random parings of oscillators which
precedes the cluster formation, occurs close to the theoretical line
predicted for a simple case of two coupled oscillators. 

\section{Many coupled oscillators - stationary states}
\label{many}
In this section we return to the dynamics of the original model
(\ref{phase_eq}),(\ref{k_eq}). As was pointed out in Section
\ref{model}, for small $\epsilon$, in the asymptotic regime 
the coupling coefficients are either non-stationary but small 
(for two oscillators operating at different frequencies) or fixed 
for two oscillators which are synchronized. So, in the limit $\epsilon\to
0$, we can neglect all $K_{ij}$ for non-synchronized oscillators, and 
define $K_{ij}=\alpha\cos(\phi_i-\phi_j)$ for synchronized oscillators. 
Substituting this into Eq.(\ref{phase_eq}), we obtain for oscillators
within a cluster,
\begin{equation}
	\dot{\phi}_i=\omega_i-\frac{\alpha}{2N}\sum_{j=1}^{N_c}\sin[2(\phi_i-\phi_j)].
	\label{ph_eq1}
\end{equation}
where unlike Ref.\cite{kuramoto} summation only applies to oscillators
within the same cluster, and $\dot{\phi}_i=\omega_i$ for oscillators
outside synchronized cluster.  It is easy to see that Eq.(\ref{ph_eq1})
is formally equivalent to the Kuramoto model for the double phases
$2\phi_i$.

\begin{figure}
\centerline{\psfig{figure=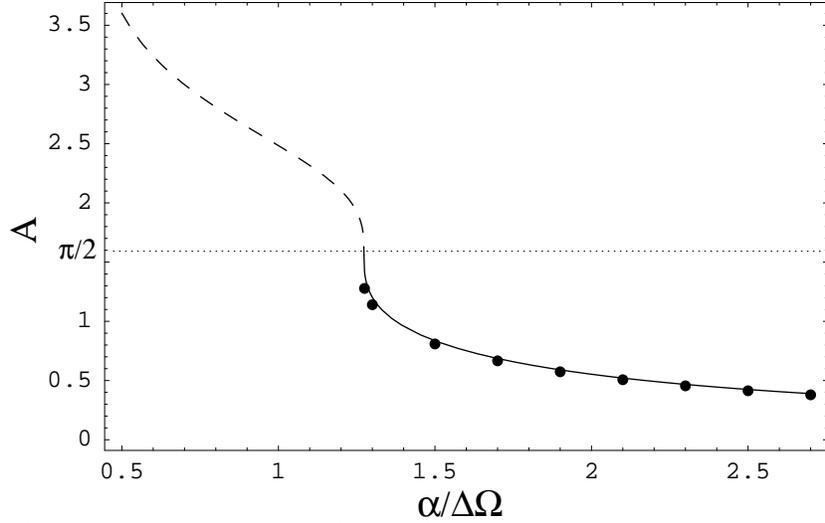,height=3in} }
\caption{Phase spread of the cluster $A$ as a function
of $\alpha/\Delta\Omega$ from Eq.(\protect\ref{A_eq}). Dots indicate the
results of numerical simulations of
Eqs.(\protect\ref{phase_eq}),(\protect\ref{k_eq}) with $50$
uniformly distributed oscillators with $\Delta\Omega=1, \epsilon=0.002$.}
\label{clusterwidth} 
\end{figure}

\begin{figure}
\centerline{\psfig{figure=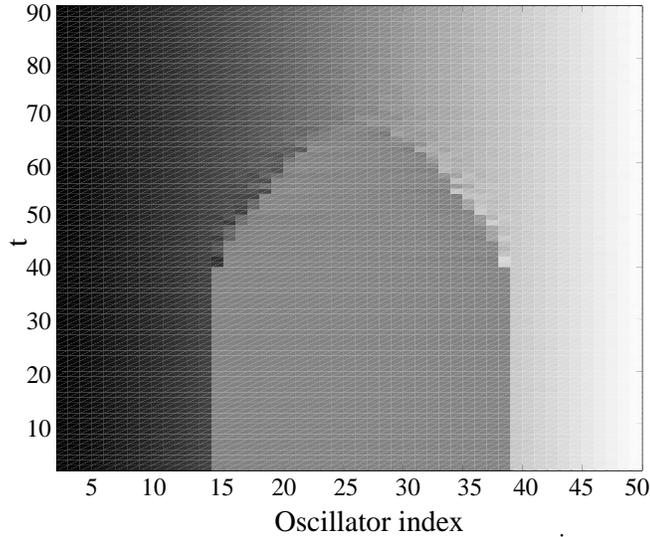,height=3in} }  
\caption{Gray-scale plot of the evolution of instantaneous oscillator
frequencies $\dot\phi_i$ for an unstable cluster at $\alpha=1.25$
Parameters of the simulation are $N=50, \epsilon=0.002, \Delta\Omega=1$.}
\label{unstableCluster} 
\end{figure}

One can investigate this model in the mean-field limit $N\to\infty$ using the 
order parameter  method \cite{kuramoto}. In this case, the complex order 
parameter reads
\begin{equation}
	\tau e^{2i\theta}=
\lim_{N \to \infty}N^{-1}\sum_{i=1}^{N_c}e^{2i\psi_i}
	=\int_{-\pi}^{\pi}n(\psi) e^{2i\psi}d\psi.
\label{op}
\end{equation}
where $\psi=\phi-\Omega t$, $\Omega$ is the frequency of the synchronized 
cluster, $n(\psi)$ is the 
phase distribution of the oscillators in the
population which belong to the cluster.
The phases of the oscillators inside the cluster are given by
\begin{equation}
\psi_i=\theta
+\frac{1}{2}\arcsin\left(\frac{2(\omega_i-\Omega)}{\alpha\tau}\right),
\label{phases}
\end{equation}
and the phase distribution $n(\psi)$ can be easily expressed via the
frequency distribution of the synchronized oscillators $g(\omega)$,
\begin{eqnarray}
n(\psi)= \left\{ \begin{array}{l}
\alpha\tau g\left( \frac{\alpha\tau}{2}\sin 2(\psi-\theta)
+\Omega \right)\cos(2(\psi-\theta))\ \mbox{for}\ B<2(\psi-\theta)<A\\
0, \ \ \mbox {outside}
\end{array}
\right.
\label{distr}
\end{eqnarray}
where $A,B$ determine size and position of the synchronized cluster,
($-\pi/2<B<A<\pi/2$). Substituting (\ref{distr}) into Eq.(\ref{op}),
we obtain the equation for the magnitude of the order parameter $\tau$
\begin{equation}
	\frac{\alpha}{2}\int_{B}^{A}g
	\left( \frac{\alpha\tau}{2}\sin x+\Omega \right)
	\cos x\ e^{ix} dx = 1.
	\label{order}
\end{equation}
The frequency spread inside the synchronized cluster is determined by
formula
\begin{equation}
\Delta\Omega_c=\frac{\alpha\tau}{2}(\sin A-\sin B),
\label{omega_c}
\end{equation}
which follows directly from Eqs.(\ref{phases}).
Thus, this system exhibit
degeneracy so that a variety of  synchronized states can be formed 
depending on initial conditions. It is particularly evident for 
the uniform distribution of oscillator
frequencies within frequency interval $\Delta\Omega$, 
$g(\omega)=\Delta\Omega^{-1}$. In this case, the cluster size drops
out, the limits of integration are symmetric, $B=-A$, and we arrive at 
\begin{equation}
        \frac{\alpha}{2\Delta\Omega}\int_{-A}^{A}
        \cos x\ e^{ix} d x= 1
        \label{order1},
\end{equation}
which reduces to an algebraic equation for phase width of the cluster $A$,
\begin{equation} \label{widthA}
A+\frac{1}{2}\sin(2A)=\frac{2\Delta\Omega}{\alpha}.
\label{A_eq}
\end{equation}
The phase width of the cluster $A$ depends only on
$\alpha/\Delta\Omega$ and does not depend on the cluster size $\tau$.  
This dependence is shown in Fig.\ref{clusterwidth}. For large $\alpha$,
$A\approx\Delta\Omega/\alpha$.
At $\alpha\to\alpha_0=4\pi^{-1}\Delta\Omega$, $A\to\pi/2$. 
The frequency spread inside the synchronized cluster is equal to
$\Delta\Omega_c=\alpha\tau\sin A$ (cf. (\ref{omega_c})).
At $\alpha<\alpha_0$, there can be no stationary synchronized clusters.
These findings are confirmed by direct numerical simulations of 
Eqs.(\ref{phase_eq}),(\ref{k_eq}). In Fig.\ref{clusterwidth} 
the phase width of the cluster is shown for several values of
$\alpha/\Delta\Omega$, and in Fig.\ref{unstableCluster}
the evolution of an unstable cluster at $\alpha=1.25<\alpha_0$ is shown.
The parameters of simulations were $N=50,\ \Delta\Omega=1,\ \epsilon=0.002$, and
the initial cluster width was chosen $\Delta\Omega_c=0.5$.

\section{Stability of the clustered state}
\label{stab}
As we have seen in the previous Section, the order parameter $\tau$
which characterizes the cluster size, can take any value from 0 to 
$\tau_{max}$ (for the uniform frequency distribution the latter 
corresponds to the whole population being
synchronized, $\tau_{max}=\Delta\Omega/\alpha\sin A$). However, this 
is true only for small enough $\epsilon$.
Just as in the case of two coupled oscillators (Sec.\ref{two}), at
sufficiently large $\epsilon> \epsilon_c$, the multistability disappears, and the
only stable state at $\alpha>\alpha_c$ is the one involving the whole
population of oscillators, $\Delta\Omega_c=\Delta\Omega$. In order to
determine $\epsilon_c$, we analyze the following
problem. Suppose we have a synchronized cluster oscillating at
frequency $\Omega$ and characterized by the order parameter
$\tau$. Let us consider a single oscillator with frequency $\omega_0$ 
adjacent to the cluster ($\omega_0=\Omega+\Delta\Omega_c/2$), and
study its dynamics neglecting its interaction with all other oscillators
and the the feedback influence of the oscillator
on the cluster. Clearly, since the frequency difference between this oscillator
and the cluster is smaller than that for any other non-synchronized
oscillator, the adjacent oscillator will be the first to get entrained 
at $\epsilon_c$. The equation for the relative phase
$\psi_0=\phi_0-\Omega t$ of this oscillator reads
\begin{equation}
 \dot{\psi_0}=\frac{\Delta\Omega_c}{2} + \frac{1}{N}\sum_{i=1}^{N_c}K_i\sin(\psi_i-\psi_0),
\label{ph_eq2}
\end{equation}
and the equation for the coupling coefficient between the oscillator and the
$i$-th member of the cluster, 
\begin{equation}
\dot{K}_i=\epsilon(\alpha\cos(\psi_i-\psi_0)-K_i).
\label{K_eq2}
\end{equation}
The summation in Eq.(\ref{ph_eq2}) is carried only over synchronized
oscillators. The phases $\psi_i$ of the synchronized oscillators are
determined by Eq.(\ref{phases}).

\begin{figure}
\centerline{ \psfig{figure=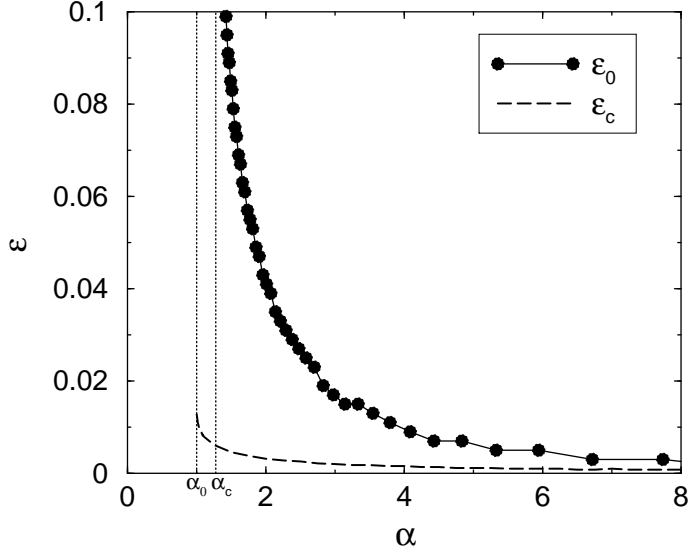,height=3.in}}
\caption{Parameter plane $(\alpha,\epsilon$). A cluster is stable at 
 $\alpha>\alpha_0$ and $\epsilon<\epsilon_0$. For $\alpha<\alpha_0$,
clusters do not exist, for $\alpha<\alpha_0$ and
$\epsilon>\epsilon_0$, cluster size increases spontaneously. Dashed line
shows the stability limit for synchronization of two neighboring
oscillators, for $\alpha>\alpha_c$ and $\epsilon>\epsilon_c$, 
random pairing of oscillators outside the main cluster occurs.
}
\label{epsilon0}
\end{figure}

In the continuum limit for the uniform frequency distribution, 
these equations become
\begin{eqnarray}
\dot{\psi_0}&=& \frac{\Delta\Omega_c}{2} +
\frac{\alpha\tau}{\Delta\Omega}
\int_{-A/2}^{A/2}K(\psi)\cos(2\psi)\sin(\psi-\psi_0)d\psi,
\label{ph_eq3}\\
\dot{K}(\psi)&=&\epsilon(\alpha\cos(\psi-\psi_0)-K(\psi)).
\label{K_eq3}
\end{eqnarray}
Eq.(\ref{ph_eq3}) can be rewritten in the form
\begin{equation}
 \dot{\psi_0}=\frac{\Delta\Omega_c}{2}
+\frac{\alpha\tau}{\Delta\Omega}(P\cos\psi_0-Q\sin\psi_0),
\label{ph_eq4}
\end{equation}
where 
\begin{eqnarray} 
P&=&\int_{-A/2}^{A/2}K(\psi)\sin \psi \cos 2\psi d\psi,\\
Q&=&\int_{-A/2}^{A/2}K(\psi)\cos \psi  \cos 2\psi d\psi.
\end{eqnarray}
Equations for $P,Q$ can be obtained using Eq.(\ref{K_eq3}),
\begin{eqnarray}
\dot{P}&=&\epsilon\left(\alpha F_-\sin\psi_0 - P\right),
\label{P_eq}\\
\dot{Q}&=&\epsilon\left(\alpha F_+ \cos\psi_0 - Q\right) ,
\label{Q_eq}
\end{eqnarray}
where$F_\pm=\frac{1}{2}\sin A \pm (\frac{1}{4}A+\frac{1}{8}\sin 2A)$.
The set of equations (\ref{ph_eq4}),(\ref{P_eq}),(\ref{Q_eq}) exhibits
the dynamics similar to Eqs.(\ref{two_osc1}) for two coupled oscillators.
At $\alpha<\alpha_0=4\pi^{-1}\Delta\Omega$, the only
attractor of the system is the solution with periodic $P$ and $Q$ and
unbounded phase $\phi$ which corresponds to ``non-entrainment''
of the oscillator by the cluster. 
For the phase variable taken {\em modulo} $2\pi$, this solution corresponds 
to a limit cycle encircling the phase cylinder. At
larger $\alpha>\alpha_0$, two fixed points appear, one of which
(stable) corresponds to the entrainment of the oscillator by the
cluster. At $\epsilon>\epsilon_0(\alpha)$, the limit cycle disappears,
and the entrained state is the only stable state of the system. Thus,
the critical line $\epsilon_0(\alpha)$ separates the region of cluster
stability with respect to entraining additional oscillators. This line
is shown in Figure \ref{epsilon0}. This line should be compared with the
critical line for pairing of two neighboring oscillators
(Fig.\ref{alpha_eps}).  Note that we have to re-scale $\tilde{\epsilon}$
and $\tilde{\alpha}$ back into original $\epsilon$ and $\alpha$ using 
$\tilde{\epsilon}=\epsilon N$, $\tilde{\alpha}=2\alpha/\Delta \Omega$. 
For large $N$, this line lies below 
$\epsilon_0(\alpha)$, which indicates that the random pairing outside
the cluster occurs before the main cluster loses its stability.
This conclusion agrees with our numerical simulations.

\section{Learning and memory}
\label{memory}
\begin{figure}[h]
\centerline{ \psfig{figure=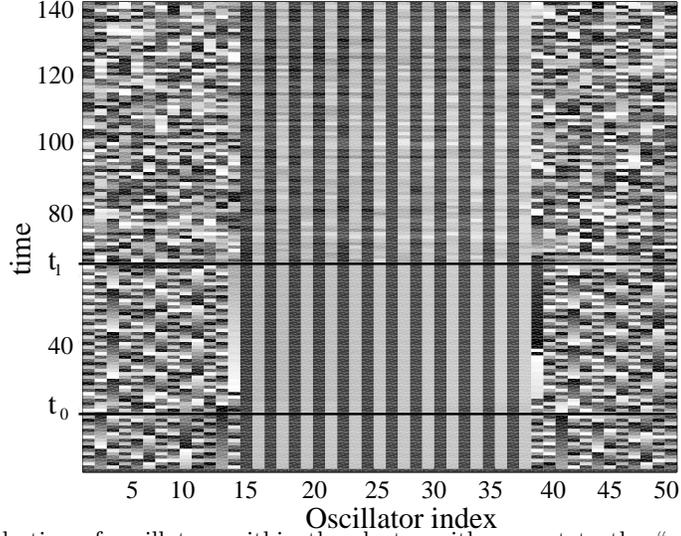,height=3.in}}
\caption{Relative phase distribution of oscillators within the cluster
with respect to the ``middle'' oscillator $i=25$. The cluster is initialized 
by external forcing of oscillators 15-37 at frequency
$\omega_0=\Omega=10.0$ for $0<t<t_0=20$. At $t>t_0$, the
external forcing was turned off, but the phase distribution within the
cluster survived. At $t>t_1=75$, oscillators were driven by random external 
fluctuation of magnitude 0.01. Despite of this forcing, the phase pattern
``memorized" by the cluster, was preserved. }
\label{clusterPhase}
\end{figure}

Information can be stored within the 
cluster in the form of relative phases of the oscillators in
the synchronized  state.  Initial learning can be accomplished via
driving a group of oscillators by external signals with identical frequency
$\omega_0$ and magnitude $K_0$, but with different phases $\phi_i^0$,
\begin{eqnarray}
\dot{\phi}_i&=&\omega_i-\frac{1}{N}\sum_{j=1}^{N}K_{ij}\sin(\phi_i-\phi_j)+
K_0\sin(\phi_i-\omega_0 t - \phi_i^0),
\label{phase_eq1} \\
\dot{K}_{ij}&=&\epsilon(\alpha\cos(\phi_i-\phi_j)-K_{ij}).
\label{k_eq1}
\end{eqnarray}
If $K_0\gg 1$, every oscillator in the group will be entrained to the
external frequency $\omega_0$. So,
the external driving signals force the oscillators to oscillate in synchrony
and, via slow synaptic dynamics Eq.(\ref{k_eq1}), form a tightly coupled 
cluster. 
If the phases $\phi_i^0$ are taken to be 0 or $\pi$, the relative phases
$\phi_i-\phi_j$ within the cluster will be also close to $0$ or $\pi$,
and so the coupling coefficients $K_{ij}$ will approach  $\pm \alpha$.
After the learning is completed after $t\approx O(\epsilon^{-1})$ time, 
the external
signal is disconnected ($K_0\to 0$). If the external frequency
$\omega_0=\Omega$, the mean frequency of oscillators within the cluster, 
the cluster remains synchronized at the same frequency, moreover, the 
phase relations among
the oscillators are effectively preserved. Even significant random
fluctuations affecting the dynamics of individual oscillators, 
do not change the robust structure of the synchronized cluster,
see Fig.\ref{clusterPhase}. 

For small $\epsilon$, after a cluster is formed, 
it becomes very stable with respect to
external noise. Furthermore, it remains stable with respect to random 
variations of their natural frequencies. We let the cluster form
similarly to described above, and then at some time $t_1$, we
perturbed the natural frequencies of all oscillators, $\bar{\omega}_i=
\omega_i+\Delta\omega \xi_i$, where $\xi_i$ are i.i.d. Gaussian random 
variable with variance 1. We can quantify the robustness
of the cluster by measuring the correlation between the
synchronized cluster structure before and after the random fluctuations
of natural frequencies have been applied. The correlation was calculated
using the formula
\begin{equation}
C=\frac{\sum_{i=1}^{N}p_i\wedge q_i}
{\sum_{i=1}^{N}p_i\vee q_i},
\label{corr}
\end{equation}
where $p_i(q_i)$ is the binary variable describing the state of the 
oscillator before(after) perturbing
the natural frequencies: $p_i(q_i)=1$ if the oscillator is entrained in
the main cluster and $p_i(q_i)=0$ otherwise. Symbols $\vee$ and $\wedge$
denote Boolean addition and multiplication, respectively. It is easy to
see that $C=1$ when cluster remains unchanged, and $C=0$ if all
oscillators leave the cluster.

\begin{figure}
\centerline{\psfig{figure=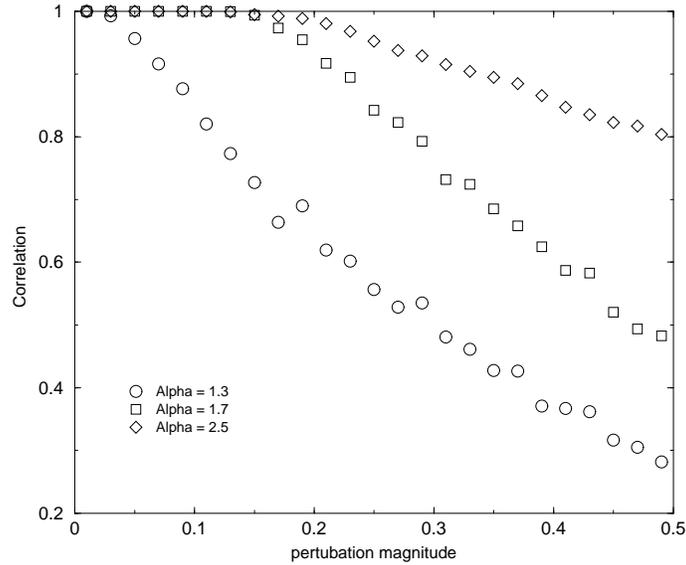,height=3in} }
\caption{Stability of cluster measured with the correlation of
the cluster before and after perturbation of natural frequencies.  The
different curves are for several values of $\alpha$, $\epsilon$ increases
the stability of the cluster only slightly and falls of parallel to the 
curves for similar $\alpha$.}
\label{correlation} 
\end{figure}

Figure \ref{correlation} 
shows correlation $C$ as a function of the perturbation magnitude
$\Delta\omega$ for several values of $\alpha$. As can be seen in the
Figure, the cluster becomes more robust with increase of $\alpha$. 

\section{Conclusion}
\label{conclusion}
In summary, we have demonstrated that a generalized Kuramoto model 
with additional equations describing slow dynamics of the coupling
matrix can be used to elucidate the underlying 
mechanism of synaptic plasticity. The slow dynamics leads to the
multi-stability:  synchronized clusters of different sizes and with
different phase relationships among oscillators 
can be stabilized. The phase differences among the oscillators
can be used as a way of storing and retrieving the information in this
system. One natural limitation of the reduced phase description of the
oscillators is that all oscillators are assumed to be in the excited
(``firing'') state.  However, in natural systems (such as biological
neural networks), some of the neurons may be in quiescent
(``non-firing'') regime. A more general description of this system
should incorporate equations for complex amplitudes of oscillators
similar to Ref.\cite{aoyagi}.

Our results may have important biological implications. In particular,
they suggest that there are fundamental reasons for the wide separation
of time scales of fast neuronal dynamics ( e.g. membrane potential
oscillations) and slow synaptic variability. They also present a very
simple (possibly, the simplest) model of the {\em synaptic reentry
reinforcement} which is believed to be the foundation of the long-term
memory stability \cite{tsien}.

We are indebted to M.I.Rabinovich and R. Huerta for fruitful discussions. 
This work was supported by the U.S. Department of Energy grant
DE-FG03-96ER14592 and by UC-MEXUS/CONACYT.


\references
\bibitem{tsien}E. Shimizu, Y.-P. Tang, C. Rampon, and J. Z. Tsien,
Science, {\bf 290}, 1170-1174 (2000).
\bibitem{abbott} L.F.Abbott and S.B.Nelson, Nature Neuroscience Suppl., 
{\bf 3}m 1178 (2000).
\bibitem{hebb} D.O. Hebb, The organization of behavior, Wiley, New York,
1949.
\bibitem{markram97} H.  Markram, J. L\"{u}bke, M. Frotscher, B. Sakmann,
Science, {\bf 275}, 213-215 (1997).
\bibitem{poo}L.I.Zhang, H.W.Tao, C.E.Holt, W.A.Harris, M.Poo, Nature,
{\bf 395}, 37 (1998).
\bibitem{hopfield}J.J.Hopfield, Proc. Natl. Acad. Sci., {\bf 79},
2554 (1982).
\bibitem{tuckwell}H. Tuckwell, {\em Introduction to Theoretical
Neurobiology.} Cambridge, 1988.
\bibitem{roberts00} P.D. Roberts, Phys. Rev. E, {\bf 62}, 4077 (2000).
\bibitem{rabinovich}H.D.I.Abarbanel, R. Huerta, M.I.Rabinovich, in
preparation. 
\bibitem{gray}C.M.Gray, J.Comput. Neurosci., {\bf 1}, 11 (1994).
\bibitem{sompolinsky}H.Sompolinsky, D.Golomb, and D.Kleinfeld, Phys.
rev. A, {\bf 43}, 6990 (1991).
\bibitem{aoyagi} T.Aoyagi, Phys. Rev. Lett., {\bf 74}, 4075 (1995).
\bibitem{hop96}F.C. Hoppensteadt and E.M.Izhikevich, Biol. Cybern., {\bf
75}, 129 (1996).
\bibitem{hop00}F.C. Hoppensteadt and E.M.Izhikevich, IEEE Trans. Neural
Networks, {\bf 11}, 734 (2000).
\bibitem{winfree}A.T.Winfree, J. Theor. Biol. {\bf 16}, 15 (1967).
\bibitem{kuramoto} Y.Kuramoto, {\em Chemical Oscillations, Waves, and
Turbulence} (Springer, New York, 1984).
\bibitem{kur_model}See review by S.H.Strogatz, Physica D {\bf 143}, 1
(2000).
\bibitem{strogatz1} M.K.Yueng and S.H. Strogatz, Phys. Rev. Lett.,
{\bf 82}, 648. (1999)
\bibitem{hop99}F.C. Hoppensteadt and E.M.Izhikevich, Phys. Rev. Lett.,
{\bf 82}, 2983 (1999).
\end{document}